\documentclass[]{aipproc}

\layoutstyle{6x9}

\usepackage{amsmath,amssymb,amsbsy,amsfonts,bm,mathrsfs}

\newcommand{\abs}[1]{\left| #1 \right|} 		
\newcommand{\ii}{\mathrm{i}}				
\newcommand{\ee}[1]{\mathrm{e}^{#1}}			
\renewcommand{\vec}[1]{\bm{#1}}				

\begin{document}

\title{A nonlinear dynamics approach to Bogoliubov excitations of Bose-Einstein condensates}


\author{M. Kreibich, H. Cartarius, J. Main, G. Wunner}{
  address={Institut f\"ur Theoretische Physik 1, Universit\"at Stuttgart, 70550 Stuttgart, Germany}
}

\begin{abstract}
We assume the macroscopic wave function of a Bose-Einstein condensate
as a superposition of Gaussian wave packets,
with time-dependent complex width parameters, insert it into
the mean-field energy functional corresponding to the Gross-Pitaevskii
equation (GPE) and apply the time-dependent variational principle. 
In this way 
the GPE is mapped onto a system of coupled equations of motion 
for the complex width parameters, which can be analyzed using the methods
of nonlinear dynamics. We perform  
a stability analysis of the fixed points of the nonlinear system, 
and demonstrate that the eigenvalues of the Jacobian  
reproduce the low-lying quantum mechanical Bogoliubov excitation 
spectrum of a condensate in an axisymmetric trap. 
\end{abstract}

\maketitle

\section{Introduction}
It is well known that at sufficiently low temperatures the 
Gross-Pitaevskii equation (GPE) \cite{Pitaevskii03a}, a nonlinear Schr\"odinger equation
for the macroscopic wave function, provides an accurate description
of the dynamics of dilute trapped Bose-Einstein condensates for
both the ground state and the excitation spectrum. For a condensate
in a trap the GPE reads, in appropriately scaled units,
\begin{equation} \label{GPE}
      \left[-\Delta + \gamma_{x}^2 x^2 + \gamma_y^2 y^2 + \gamma_z^2 z^2 + 8\pi a
        |{\psi(\vec{r}, t)}|^2 \right] \psi(\vec{r}, t) = {\mathrm i}
      \frac{\partial}{\partial t} \psi(\vec{r}, t) .
    \end{equation}
Here $a$ is the s-wave scattering length corresponding to the
short-range contact interaction between the condensate atoms, and
$\gamma_{x}, \gamma_{y}, \gamma_z$ are the frequencies of the traps confining
the condensate in the three spatial directions. 
Because of its
nonlinearity, the GPE can in general only be solved numerically,
e.g. by imaginary time evolution (cf., e.g., \cite{Koeberle12a}).

A full-fledged alternative to numerical quantum calculations
is a variational approach \cite{Heller81a} in which the trial functions are
superpositions of $N$ different Gaussians
\[\Psi(\textbf{r}) = \sum_{k = 1}^{N} {\mathrm e}^{{\mathrm i}\left(a_x^{k}(t)~x^{2}+a_y^{k}(t)~y^{2}+a_z^{k}(t)~z^{2}+\gamma^{k}(t) \right)}
\equiv \sum_{k = 1}^{N}g^{k}(\textbf{y}^k(t),\textbf{r})
		\]
with $3N$  complex width parameter functions  $\vec{a}^k(t)$
and $N$ functions
$\gamma^k(t)$ which give the  weights and phases of the individual 
Gaussians.  Applying the Dirac-Frenkel variational principle
\cite{Dirac1930a, McLachlan1964a}, i.e. requiring
$||{\rm i} \phi(t) -H \Psi(t)||^2$  to be minimum with respect  to the
choice of $\phi$, and afterwards setting 
$\phi = \dot \psi$, leads to equations of motion for the variational 
parameters which formally can be written
\begin{equation} \label{vareq}\textstyle{  
K \dot {\vec \lambda} = -{\mathrm i} {\vec h} \hbox{~with~} K = \left\langle \frac{\partial \psi}{ \partial{{\vec \lambda}}}
 \Big|\frac{\partial \psi}{\partial{{\vec \lambda}}} \right\rangle, 
  {\vec h} = \left\langle \frac{\partial \psi}
 { \partial{{\vec \lambda}}}
 \Big|H \Big|\psi \right\rangle} ,
\end{equation}
where $\vec \lambda$ stands for all variational parameters. 
Stationary solutions are then the fixed points of the equations
of motion. 

This method has been successfully applied to the description
of Bose-Einstein condensates. 
Even for condensates with long-range
interactions (monopolar and dipolar), in addition to the short-range 
contact interaction, the method leads to highly accurate results
for energies and wave functions \cite{Rau10a, Rau10b, Rau10c}. 
For example it well
reproduces structured ground states that can appear in condensates
of dipolar atoms \cite{lahaye09b} close to collapse, and which first had been 
predicted numerically \cite{Wilson09a}. 
The method has also successfully been applied to the dynamics of anisotropic
solitons in dipolar Bose-Einstein condensates \cite{eichler11}, and  interacting multi-layer
stacks of such condensates \cite{Junginger10a}.
Moreover, the method is capable of giving access to regions of the space of solutions of the GPE
that are difficult or impossible to investigate by conventional numerical 
calculations. In this way it was possible to reveal phenomena characteristic of nonlinear classical
systems in the theory of Bose-Einstein condensates 
such as the appearance of  bifurcations \cite{Rau10c}, 
exceptional points \cite{Cartarius09b}, and
the transition from order to chaos \cite{Koeberle09a}.

In this paper we will address the two different concepts of stability
that are involved in the two approaches. In the full 
quantum mechanical treatment the stability of Bose-Einstein condensates   
is investigated by calculating Bogoliubov excitations of the ground state.
In the nonlinear dynamics approach stability is
determined by calculating the eigenvalues of the Jacobian at the fixed
points. Is there a relation between these eigenvalues and the frequencies of 
the quantum mechanical Bogoliubov excitations?

\section{Bogoliubov-de Gennes equations}

We consider a Bose-Einstein condensate in an axisymmetric trap,
i.e. in (\ref{GPE}) only two trap frequencies, $\gamma_z$, $\gamma_{\rho}$
appear. If $\psi_0(\mathbf{r})$ is the stationary solution of the GPE,
with $\mu$ its chemical potential,
the Bogoliubov ansatz for elementary excitations is
\begin{equation}
 \psi(\mathbf{r}, t) = \left[
    \psi_0(\mathbf{r}) + u(\mathbf{r}) \ee{-\ii\omega t} +
    v^*(\mathbf{r}) \ee{\ii\omega t} \right] \ee{-\ii\mu t} .
\end{equation}
Inserting this into (\ref{GPE}) and linearizing in the perturbations
$u$ and $v$ around the stationary solution yields the Bogoliubov-de Gennes
(BDG) equations
 \begin{eqnarray}\label{BDGE}
      \omega u (\mathbf{r}) &= \left[-\Delta + \gamma_{\rho}^2 \rho^2
        + \gamma_z^2 z^2 - \mu + 16 \pi a |\psi_0 (\mathbf{r})|^2
      \right]
      u(\mathbf{r}) + 8\pi a \left( \psi_0 (\mathbf{r}) \right) ^2 v
      (\mathbf{r})  , \\
      -\omega v (\mathbf{r}) &= \left[-\Delta + \gamma_{\rho}^2 \rho^2
        + \gamma_z^2 z^2 - \mu + 16 \pi a |\psi_0 (\mathbf{r})|^2
      \right] v (\mathbf{r}) + 8 \pi a
      \left( \psi_0^* (\mathbf{r})\right)^2 u (\mathbf{r}).
    \end{eqnarray}
Because of the axial symmetry we can make the separation ansatz
$u = \ee{\ii m \varphi}u_m(\rho,z)$, $v = \ee{\ii m \varphi}v_m(\rho,z)$,
where $m = 0, 1, 2, \dots$ is the azimuthal (angular momentum)
quantum number of the perturbation. Since the system is also invariant
under reflections $z \to -z$, the excitation modes can also be classified
according to their $z$-parity, even or odd. The equations were solved 
following a procedure described by Ronen et al. \cite{Ronen06a},
which  takes advantage of discrete 
Hankel-Fourier transforms to move between space and momentum
space representations and uses the Arnoldi method to efficiently compute 
the low lying eigenvalues.

\begin{figure}\label{fig1}
  \includegraphics[height=.3\textheight]{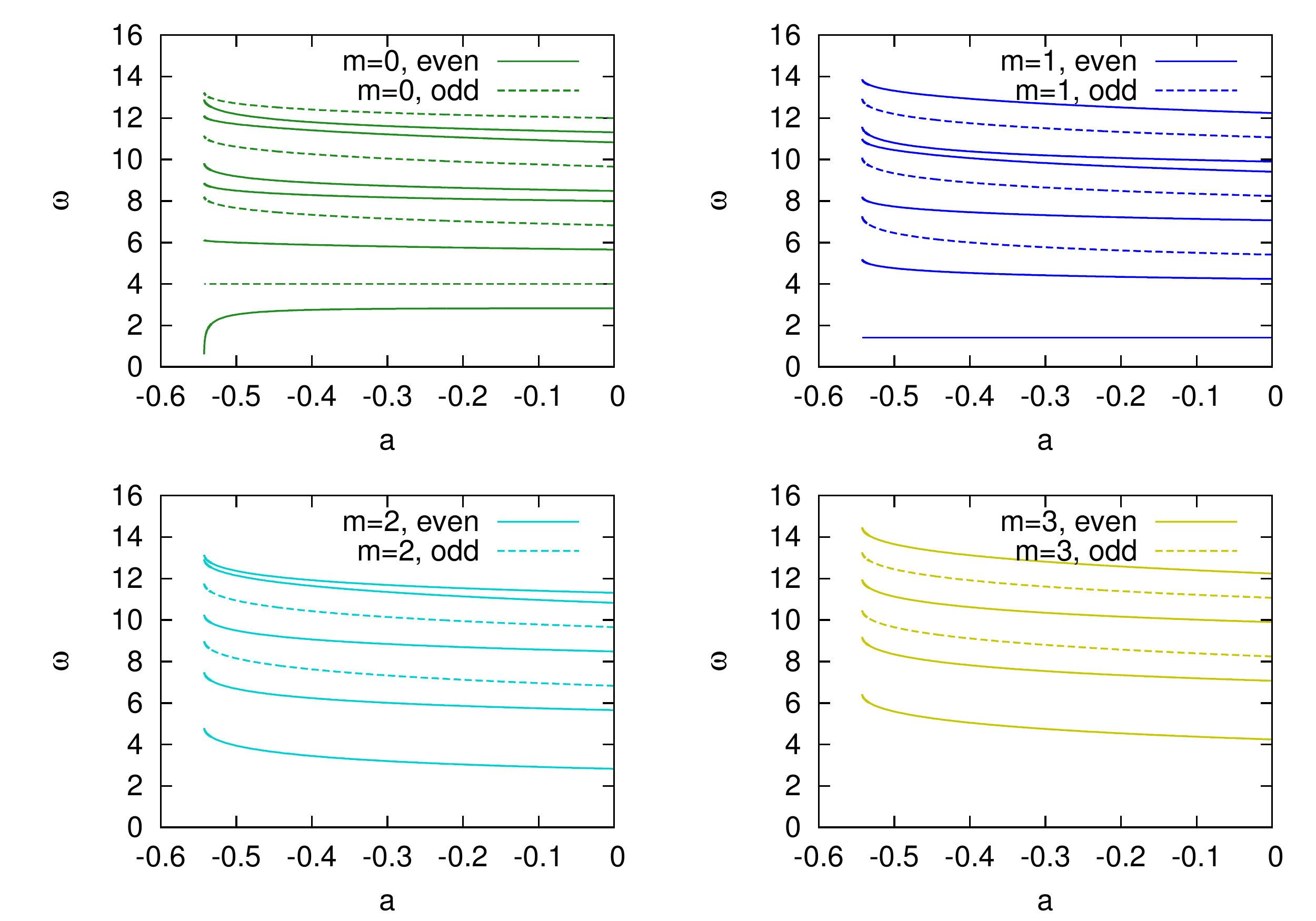}
  \caption{Frequencies of Bogoliubov excitations of the ground state
of a BEC in an axisymmetric trap as functions of the scattering length
for different values of the azimuthal quantum number $m$. 
The lowest lying modes are shown for each $m$. The trap frequencies
chosen are $\gamma_{\rho} = 1/\sqrt{2}$, $\gamma_z = 2$.
}
\end{figure}

Figure~\ref{fig1} shows the results for the frequencies of low lying 
modes as functions of the scattering length for angular excitations with
$m=0, 1, 2$ and $3$ for a pancake-shaped trap with  $\gamma_{\rho} = 1/\sqrt{2}$, $\gamma_z = 2$. It can be seen that all eigenfrequencies are real
over a wide range of the scattering length, which just confirms 
that the ground state is stable with respect to elementary
excitations. At $a \approx -0.55$ the frequency of the lowest $m=0$ mode
turns negative, i.e. it is this mode which induces the collapse of
the condensate at this scattering length.

\section{Coupled Gaussian Wave Packets }
 The variational ansatz for the wave function that respects the remaining 
symmetries of the systems,
  axial symmetry and $z$ parity, is
  \begin{equation}\label{var}
    \psi = 
    \sum\limits_{m} \ee{\ii m \phi} \rho^{\abs{m}}  \sum\limits_{p=0,1}z^p \left(\sum\limits_{k=1}^N    
    d_{m,p}^k  \ee{-(A_\rho^k \rho^2
      + A_z^k z^2 + p_z^k z + \gamma^k)}\right),
  \end{equation}
  with
\begin{equation*}
 A_\rho^k = A_\rho^k(t), ~A_z^k = A_z^k(t), ~d_{m,p}^k = d_{m,p}^k(t) \in 
{\mathbb{C}}  . 
\end{equation*}
The factor $z^p$ distinguishes states with even or odd $z$ parity, $p = 0, 1$. 
For a fixed value of $m$ we insert this ansatz into the time-dependent 
variational principle and derive the equations of motion
for the variational parameters. The advantage of using Gaussians is that in
calculating the energy functional all necessary integrals can be expressed
analytically. The fixed points are then determined by a nonlinear root-search
of the equations of motion. Finally the Jacobian is evaluated at the 
fixed points and diagonalized.

\begin{figure}[t]\label{fig2}
\begin{minipage}[b]{0.5\linewidth}
\centering
\includegraphics[scale=0.285]{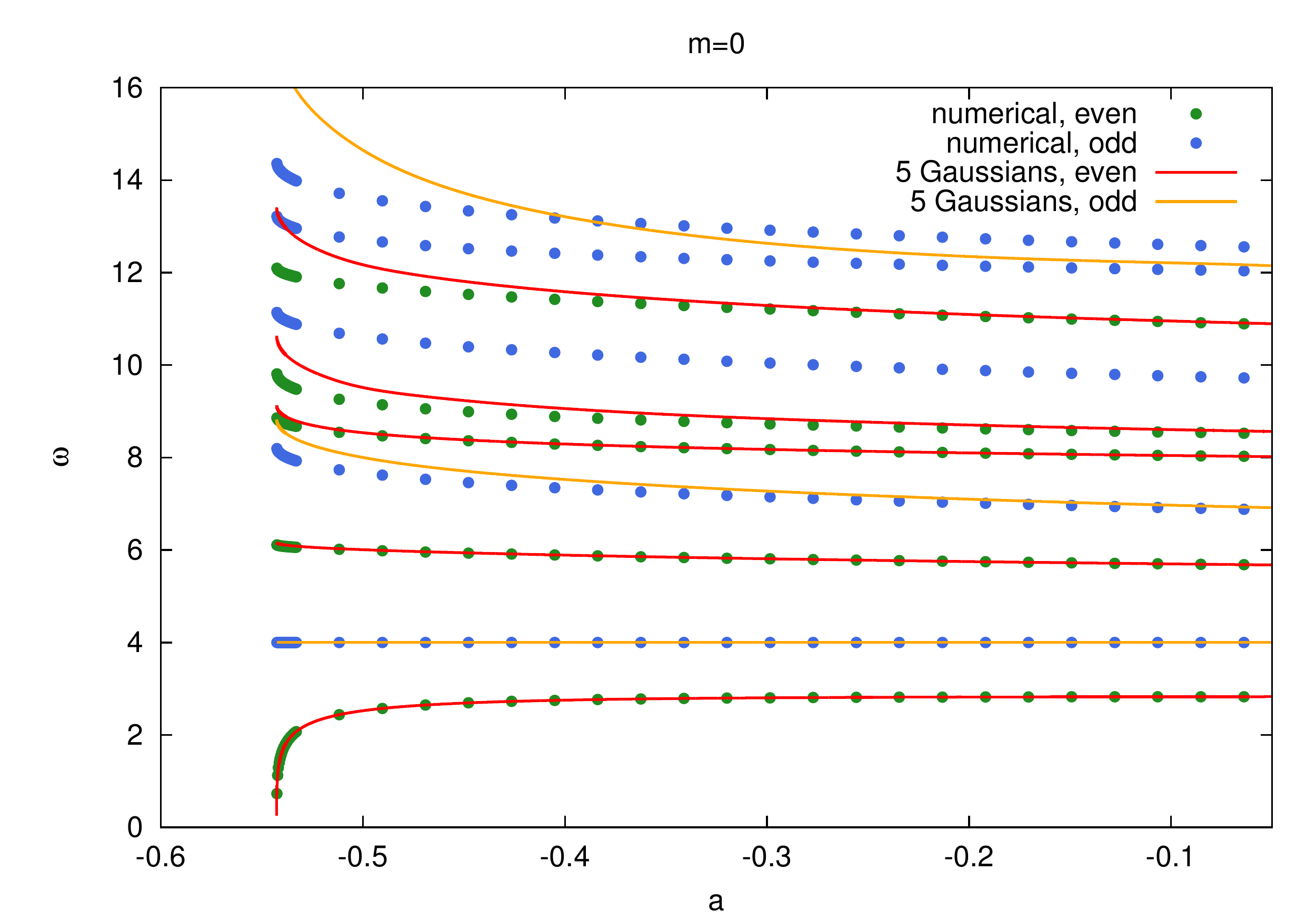}\\
\includegraphics[scale=0.285]{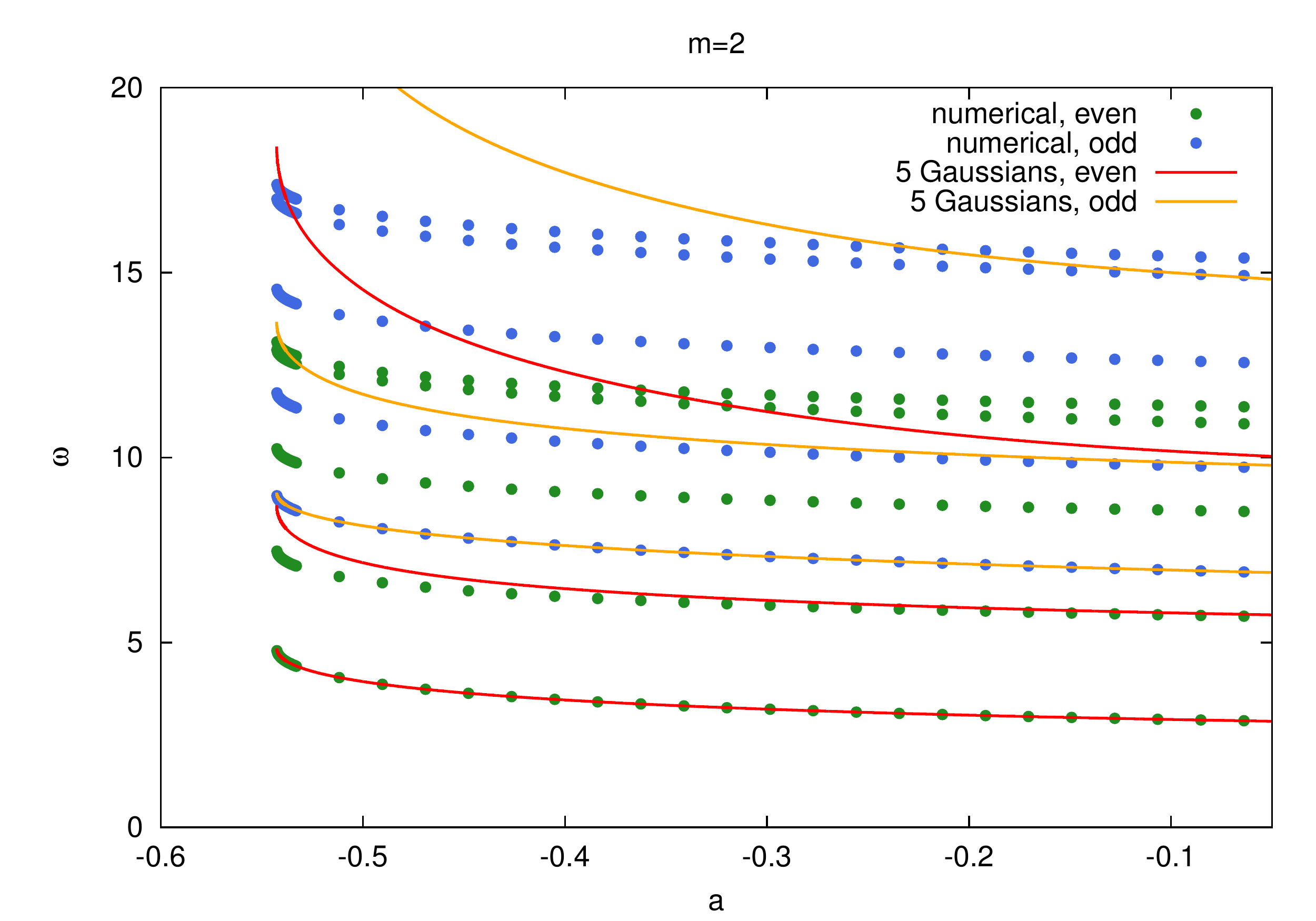}
\caption{default}
\end{minipage}
\hspace{0.0cm}
\begin{minipage}[b]{0.5\linewidth}
\centering
\includegraphics[scale=0.285]{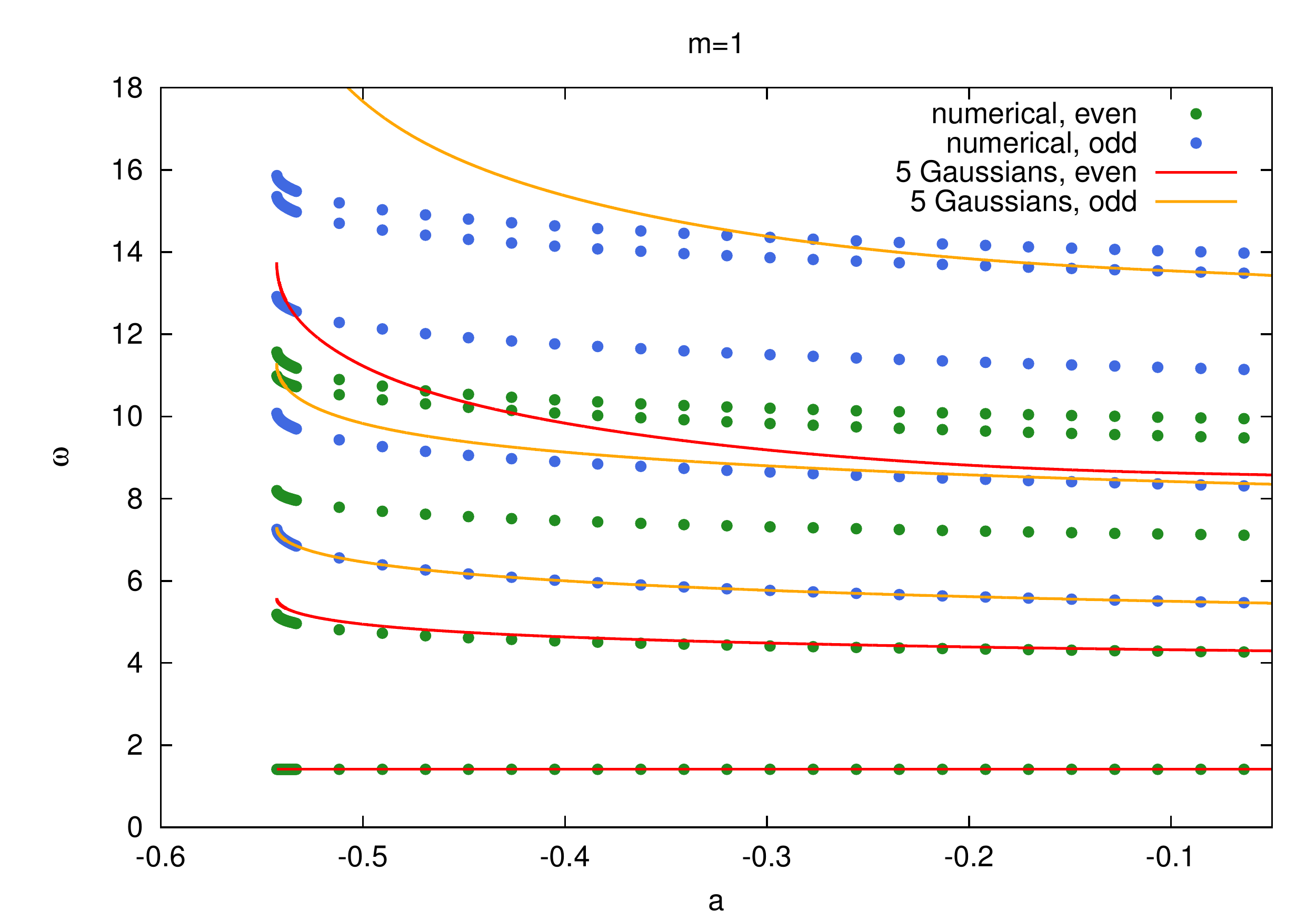}\\
\includegraphics[scale=0.285]{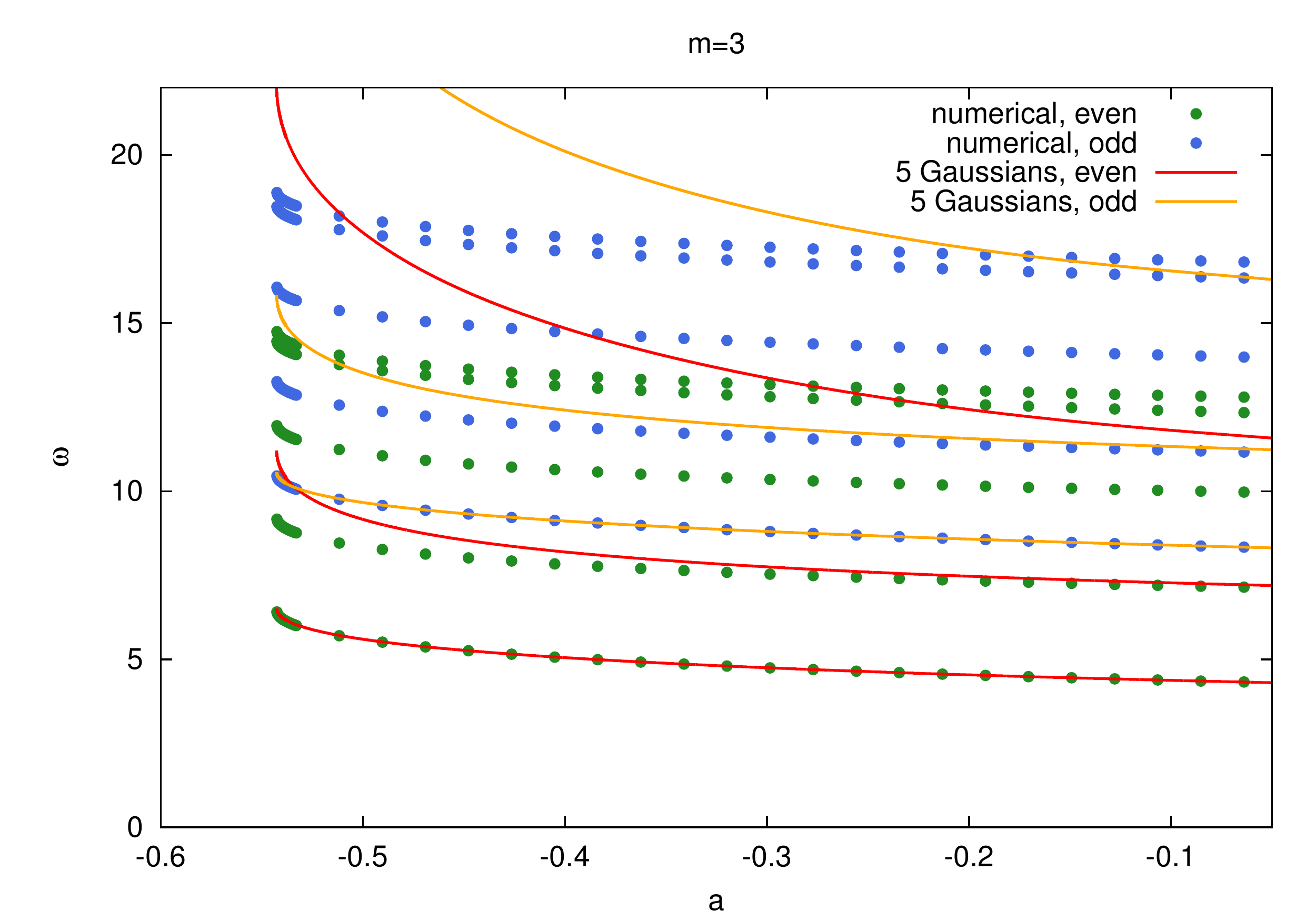}
\caption{default}
\end{minipage}
\caption{Comparison of the full-numerical Bogoliubov
    spectrum of a BEC in an axisymmetric trap 
with attractive contact interaction
    with the spectrum
    obtained from the variational ansatz with 5 coupled Gaussians
for the azimuthal excitations with $m = 0, 1, 2, 3$. As in Fig.~\ref{fig1} the trap frequencies
chosen are $\gamma_{\rho} = 1/\sqrt{2}$, $\gamma_z = 2$. }   
\end{figure}
The results are shown in
Fig.~\ref{fig2} where we 
compare the eigenvalue spectrum of the Jacobian calculated
in dependence on the scattering length with the
frequencies of the corresponding Bogoliubov excitations. One
recognizes that in particular the eigenvalues of the lowest modes of
each  azimuthal excitation agree well with the eigenfrequencies
of the Bogoliubov excitations. It is only close to the critical
scattering length that small deviations appear.  For
the higher modes with eigenvalues of the Jacobian $\omega > 10$, only far away
from the collapse point the variational and full-numerical results
still approximately correspond to each other, and in the vicinity of the
critical scattering length the Jacobi eigenvalues can reproduce the
behavior of the Bogoliubov excitation eigenfrequencies only qualitatively.

\begin{figure}\label{fig3}
\includegraphics[scale=0.45]{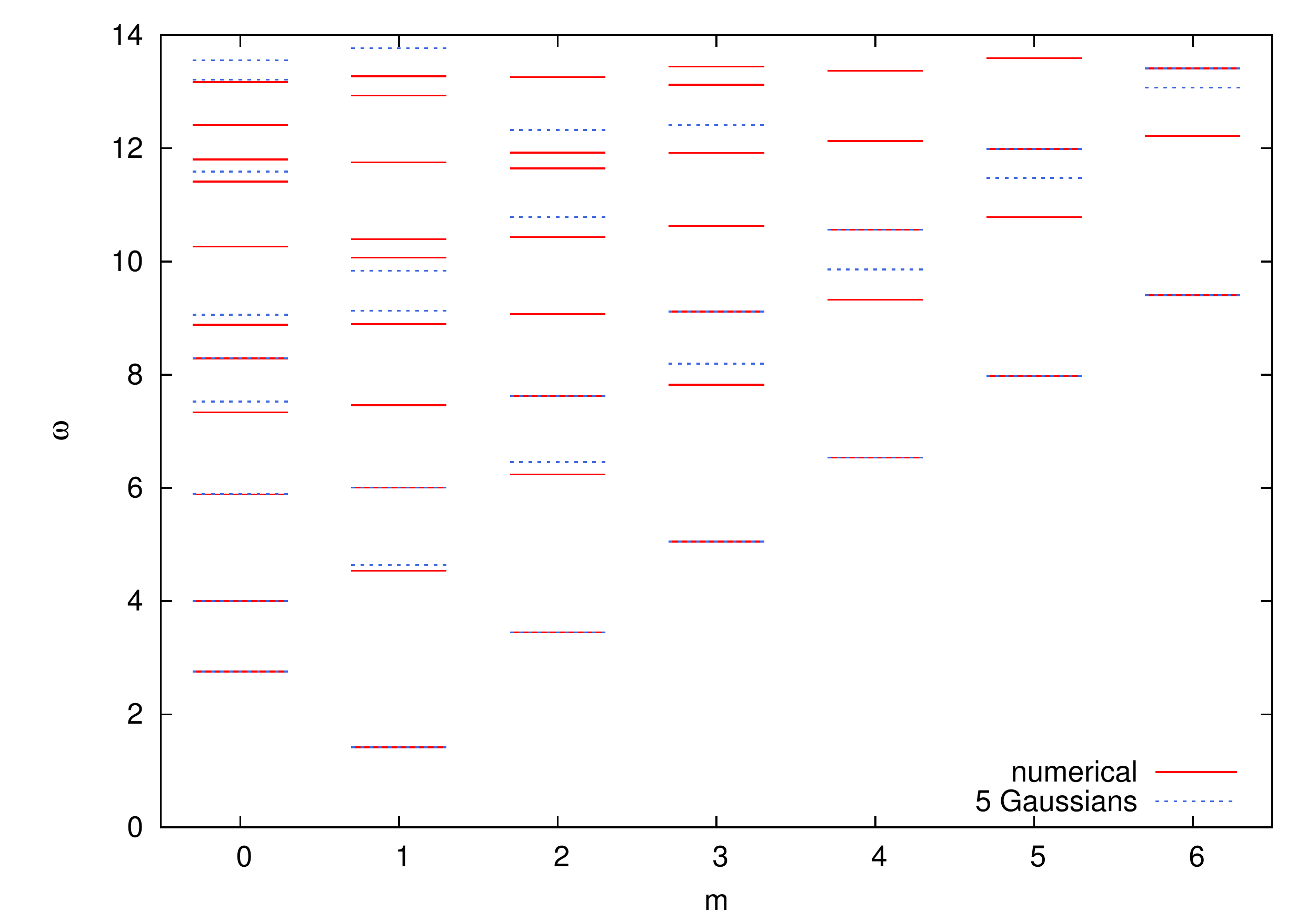}
\caption{ Comparison of both spectra as in
    Fig.~\ref{fig2}, but here for a
    fixed scattering length of $a=-0.4$ and azimuthal excitations up to
    $m=6$.  For $m=0$ the variational ansatz reproduces the Bogoliubov
 frequencies very well for the four lowest  modes, and  with only small 
deviations for the two lowest modes in the 
 bands with $m >0$.
}
\end{figure}

We have  also applied  the variational ansatz to azimuthal excitations 
up to $m=6$. The results for a fixed scattering
length of $a=-0.4$ are presented in
Fig.~\ref{fig3}. One recognizes a very
good agreement for the lowest modes in each $m$ band, and small
differences for the second-lowest modes. 

\section{Conclusions and Outlook}                                       

We have demonstrated that for
condensates with attractive short-range interaction in an axisymmetric 
trap the eigenvalues of
the Jacobian matrix calculated at the fixed point corresponding
to the ground state in the 
variational ansatz (\ref{var}) with coupled Gaussian wave packets
{\em quantitatively} coincide with the
eigenfrequencies of the lowest quantum mechanical Bogoliubov modes. 

This is remarkable because
it establishes a link between 
two completely different concepts
of stability: On the one hand the quantum mechanical stability
of a wave function with respect to elementary excitations, 
on the other hand the stability
of a classical dynamical system at a fixed point with respect to
small perturbations.

This finding is not restricted to condensates in axisymmetric traps. 
Kreibich et al. \cite{Kreibich12a} have demonstrated that for 
condensates in radially symmetric trapping potentials 
there is also a good agreement between the quantum mechanical eigenfrequencies
of the lowest Bogoliubov excitations and the eigenvalues of the Jacobian 
stability matrix. Their analysis is more involved since, to account for the spherical symmetry of the 
systems, the variational ansatz (\ref{var}) has to be modified
to include also spherical harmonics
 \begin{equation}
  \label{eq:varansatzsph}
  \psi = \sum\limits_{k=1}^N
  \left(
    1 +
    \sum\limits_{(l,m) \neq (0,0)} d_{lm}^k Y_{lm} (\theta,\phi) r^l
  \right)
  \ee{-A_r^k r^2 - \gamma^k}.
\end{equation}
The amplitudes $d_{lm}^k$ of the spherical harmonics are additional 
variational functions of time, whose equations of motion must be
obtained from the time-dependent 
variational principle, together with those for the variational functions
entering the Gaussians. The fixed points again correspond to the 
stationary ground states.

The variational approach should also be extended to Bose condensates in which the
atoms interact via long-range forces. The most prominent
examples are condensates of atoms with a large magnetic moment
such as $^{52}$Cr \cite{Griesmaier05a}, $^{164}$Dy \cite{Lu10a,Lu11a}, and other lanthanides \cite{McLeland06a}, in which the
dipole-dipole interaction is active. 

As a model system with long-range
interaction, monopolar  
condensates, with {\em gravity-like} attractive long-range interaction, 
have been proposed \cite{ODell00a}. The interaction is induced by shining
an appropriate arrangement of lasers on the condensate. 
Such condensates are unique in that
they  exhibit the phenomenon of self-trapping, without an external potential.
They are of special theoretical
interest since their investigation can serve as a useful guide
to studies of the more complicated situation of Bose-Einstein 
condensates with the anisotropic dipole-dipole interaction. 
For example, the occurrence of exceptional points in Bose-Einstein 
condensates was first shown for the monopolar 
interaction \cite{Cartarius08a} before it was also demonstrated to appear
in dipolar condensates \cite{Gutoehrlein12a}. 

Kreibich et al. \cite{Kreibich12a} have already looked at Bose-Einstein condensates with $1/r$ interaction in radially symmetric traps. They discovered that
for self-trapped condensates a good agreement between
the eigenvalues of the Jacobian and the eigenfrequencies of 
Bogoliubov excitations is present only for 
the very lowest modes, while the variational approach works less well
for higher modes.  The reason is  that for condensates in a trap the
confining radially symmetric harmonic potential 
dominates the properties of the system over a wide range of the
scattering length, and the 
interactions quasi act as perturbations.  Therefore, a
variational ansatz in which the radial part is determined by
Gaussians is very well adapted to describe the stationary solutions and
their excitations.

On the other hand, for the special situation of a self-trapped 
monopolar condensate, the interactions alone 
determine the
properties of the system. Asymptotically, for $r \to
\infty$ the BDG equations  assume the form of the Schr\"odinger
equation of the hydrogen atom. Therefore for large $r$ the solutions
$u$ and $v$ assume the decaying shape of hydrogen wave functions  
$\propto \exp(-\alpha r)$, with some $\alpha > 0$.  A
variational ansatz with coupled Gaussians and spherical harmonics obviously is
not well suited to reproduce this asymptotic behavior. However, as
soon as a radially symmetric trap is switched on, the agreement between
the quantum mechanical and the nonlinear dynamics excitations 
is present again also for the higher modes. We can therefore
conclude that this agreement is of generic type, and should also
be present for dipolar condensates. Investigations along this line
are in progress.

\begin{theacknowledgments}
This work has been supported by Deutsche Forschungsgemeinschaft
under contract Ma 1639/10-1.
\end{theacknowledgments}

\bibliographystyle{aipproc}   


\end{document}